\documentstyle[a4,11pt]{article}
\begin{document}

\begin{titlepage}
\vskip 2cm
\begin{flushright}
Preprint CNLP-1994-03
\end{flushright}
\vskip 2cm
\begin{center}
{\large {\bf A (2+1)-dimensional integrable spin model ( the M-XXII
equation) and Differential geometry of curves/surfaces }}\footnote{Preprint
CNLP-1994-03. Alma-Ata. 1994 \\
cnlpmyra@satsun.sci.kz}
\vskip 2cm

{\bf R. Myrzakulov }

\end{center}
\vskip 1cm
Centre for Nonlinear Problems, PO Box 30, 480035, Alma-Ata-35, Kazakhstan

\vskip 1cm

\begin{abstract}
Using the differential geometry of curves and surfaces the Lakshmanan
equivalent counterpart of the M-XXII equation is found. It is shown
that these equations are too gauge eqivalent to each other. Also the
gauge equivalence between the Strachan and M-XXII$_{q}$
equations is established. Some integrals of motion are presented. It is
well known that the spectral parameter of the some (2+1)-dimensional
soliton equations satisfies the following equations: $\lambda_{t} =
\kappa \lambda^{n}\lambda_{y}$ (a nonisospectral problems). We present
the simplest exact solutions of this equation.
\end{abstract}


\end{titlepage}

\setcounter{page}{1}

\newpage

\begin{center}
{\bf 1. Introduction}
\end{center}

Dynamics of numerous nonlinear phenomena can be modelled by
the nonlinear partial differential equations(NPDE), which describe an
evolution of curves and surfaces[1-6].  A remarkable subclass of
NPDE are soliton equations. The soliton theory as a new branch
of mathematical physics, has been developed during the
last three decades.
Solitons has found an enormous variety of applications in various
branches of science: biology, chemistry, physics, geometry and so on[7].
Recently many efforts have been made to generalize the soliton
theory to the (2+1)-dimensional case[8-11]. In contrast with
the (1+1)-dimensional case, the (2+1)-dimensional soliton
equations display a richer phenomology.

Integrable spin systems, besides being mathematically beautiful,
have the important physical applications. It is well known
that in the study of NPDE, the gauge[12] and Lakshmanan[3,13] equivalences
take place between the some NPDE. Recently a new class integrable and
nonintegrable spin systems including a multidimensional systems
and the (2+1)-dimensional version of Lakshmanan equivalence were
presented[14,18,19]. In this letter we will find the Lakshmanan and gauge equivalent
counterpart of one of these systems - the Myrzakulov XXII(M-XXII)
equation.

\begin{center}
{\bf 2. The Myrzakulov XXII equation}
\end{center}

Consider the M-XXII equation[14]
$$
-iS_t=\frac{1}{2}([S,S_y]+2iuS)_x+\frac{i}{2}V_1S_x-2ib^2 S_y \eqno(1a)
$$
$$
u_x=-\vec S(\vec S_x\wedge \vec S_y), \quad
V_{1x}=\frac{1}{4b^2}(\vec S^2_x)_y. \eqno(1b)
$$
where $\vec S = (S_{1}, S_{2}, S_{3})$ is a spin vector,
$\vec S^{2} = E = \pm 1$.
These equations are integrable. The corresponding Lax representation is
given by[14]
$$
\Phi_{x} = \{ - i(\lambda^{2} - b^{2})S + \frac{\lambda - b}{2b}SS_{x}\}\Phi  \eqno(2a)
$$
$$
\Phi_{t} = 2\lambda^{2}\Phi_{y} + \{(\lambda^{2} - b^{2})(2A + B) +
(\lambda - b)C\}\Phi  \eqno(2b)
$$
with
$$
A = \frac{1}{4}([S,S_{y}] +2iuS) + \frac{i}{4}V_{1}S,\quad
B = \frac{i}{2}V_{1}S, \quad
C = - \frac{V_{1}}{4b^{2}}SS_{x} +\frac{i}{2b}\{ S_{xy} - [S_{x}, A]\}.
$$
Here
spin
matrix has the form
$$
S= \pmatrix{
S_3 & rS^- \cr
rS^+ & -S_3
}, \quad S^2=I,\quad r^2=\pm1,\quad S^{\pm } = S_{1} \pm i S_{2}.
$$

\begin{center}
{\bf 3. The Lakshmanan equivalent counterpart of the Myrzakulov
XXII equation}
\end{center}

Now find the Lakshmanan equivalent counterpart of the M-XXII equation (1)
for the case $E = +1$ (for the case $E = -1$ see, e.g., [14]).
To this end  we use the two geometrical approaches(D- and C-approaches).

a){\it The D-approach}. Consider
a space $R^{3}$ with an orthonormal basis
$\vec e_{i},  i=1,2,3$. In this space consider the motion of curves. Model of this
curves are given by[14]
$$
\left ( \begin{array}{ccc}
\vec e_{1} \\
\vec e_{2} \\
\vec e_{3}
\end{array} \right)_{x} = C
\left ( \begin{array}{ccc}
\vec e_{1} \\
\vec e_{2} \\
\vec e_{3}
\end{array} \right),\quad
\left ( \begin{array}{ccc}
\vec e_{1} \\
\vec e_{2} \\
\vec e_{3}
\end{array} \right)_{y} = D
\left ( \begin{array}{ccc}
\vec e_{1} \\
\vec e_{2} \\
\vec e_{3}
\end{array} \right), \quad
\left ( \begin{array}{ccc}
\vec e_{1} \\
\vec e_{2} \\
\vec e_{3}
\end{array} \right)_{t} = G
\left ( \begin{array}{ccc}
\vec e_{1} \\
\vec e_{2} \\
\vec e_{3}
\end{array} \right) \eqno(3)
$$
with
$$
C =
\left ( \begin{array}{ccc}
0   & k     & 0 \\
-Ek & 0     & \tau  \\
0   & -\tau & 0
\end{array} \right) ,\quad
D =
\left ( \begin{array}{ccc}
0       & m_{3}  & -m_{2} \\
-Em_{3} & 0      & m_{1} \\
Em_{2}  & -m_{1} & 0
\end{array} \right),\quad
G =
\left ( \begin{array}{ccc}
0       & \omega_{3}  & -\omega_{2} \\
-E\omega_{3} & 0      & \omega_{1} \\
E\omega_{2}  & -\omega_{1} & 0
\end{array} \right).
$$
Hence we have
$$C_y - D_x + [C, D] = 0,\quad
C_t - G_x + [C, G] = 0,\quad
D_t - G_y + [D, G] = 0.  \eqno (4)
$$
From the first equation  we obtain
$$ (m_{1}, m_{2}, m_{3}) =
(\partial ^{-1}_{x}(\tau_{y} + k m_{2}), m_{2},
\partial ^{-1}_{x}(k_{y} - \tau m_{2}) ),\quad
m_{2} = \partial ^{-1}_{x}(\tau m_{3} - k m_{1}) \eqno(5) $$

Now let $\vec S = \vec e_{1}$. Then from equations(1) and(3) we get
$\omega_{j}$.
Let us now introduce the new function $q$ by
$$
q = \frac{k}{2b}\exp{i[\frac{1}{8}\partial^{-1}_{x}(k^{2}b^{-2} - 4\tau)
- 2b^{2}x]}  \eqno(6)
$$
Then the function $q$ satisfies the following equations[14]
$$
iq_t + q_{yx} + \frac{i}{2}[(V_1q)_x - V_2 q - qpq_y] = 0 \eqno(7a)
$$
$$
ip_t - p_{yx} + \frac{i}{2}[(V_1p)_x + V_{2}p - qpp_y] = 0 \eqno(7b) $$
$$ V_{1x}=(pq)_y, \quad  V_{2x}=p_{yx}q-pq_{yx} \eqno(7c)
$$
where $p = E\bar q$. This  set of equations is the
L-equivalent counterpart
of the M-XXII equation(1). As it seems to us, equations (7) are new.
We will call (7) the M-XXII$_{q}$ equation.

b) {\it The C-approach}. Note that the L-equivalent counterpart of equation(1)
we can find also using the surface
approach, e.g., the C-approach[14]. Let us show it.
Consider the motion of surface in
the 3-dimensional space which generated by a position vector
$\vec r(x,y,t) = \vec r(x^{1}, x^{2}, t)$. According to the C-approach[14],
$x$ and $y$ are local coordinates on the surface. The
first and second fundamental forms in the usual notation are given by
$$ I=d\vec r d\vec r=Edx^2+2Fdxdy+Gdy^2, \quad
II=-d\vec r d\vec n=Ldx^2+2Mdxdy+Ndy^2 \eqno (8) $$
where $$ E=\vec r_x\vec r_x=g_{11},\quad  F=\vec r_x\vec r_y=g_{12},\quad
G=\vec r_{y^2}=g_{22},   $$
$$ L=\vec n\vec r_{xx}=b_{11},\quad M=\vec n\vec r_{xy}=b_{12},\quad
N=\vec n\vec r_{yy}=b_{22},\quad \vec n=\frac{(\vec r_x \wedge \vec r_y)}.
{|\vec r_x \wedge \vec r_y|}  $$

In this case, the set of equations of the C-approach[14],
becomes
$$ \vec r_{t} = W_{1}\vec r_x + W_{2} \vec r_y + W_{3} \vec n \eqno (9a) $$
$$ \vec r_{xx}=\Gamma^1_{11} \vec r_x + \Gamma^2_{11} \vec r_y +L \vec n \eqno (9b) $$
$$ \vec r_{xy}=\Gamma^1_{12} \vec r_x + \Gamma^2_{12} \vec r_y +M \vec n \eqno (9c)$$
$$ \vec r_{yy}=\Gamma^1_{22} \vec r_x + \Gamma^2_{22} \vec r_y + N \vec n \eqno (9d)$$
$$ \vec n_x=p_1 \vec r_x+p_2 \vec r_y \eqno (9e) $$
$$ \vec n_y=q_1 \vec r_x+q_2 \vec r_y \eqno (9f) $$
where $W_{j}$ are some functions, $ \Gamma^k_{ij} $ are the Christoffel symbols of the second kind defined by
the metric $ g_{ij} $ and $ g^{ij}=(g_{ij})^{-1} $ as
$$ \Gamma^k_{ij}=\frac{1}{2} g^{kl}(\frac{\partial g_{lj}} {\partial x^i}+
   \frac {\partial g_{il}}{ \partial x^j}-\frac{\partial g_{ij}}
   {\partial x^l}) \eqno (10) $$
The coefficients $ p_i, q_i $ are given by
$$ p_i=-b_{1j}g^{ji}, \,\,\, q_i=-b_{2j}g^{ji} \eqno (11) $$
The compatibility conditions $ \vec r_{xxy}=\vec r_{xyx} $ and
$ \vec r_{yyx}=\vec r_{xyy} $ yield the following Mainardi-Peterson-Codazzi
equations (MPCE)
$$ R^l_{ijk} = b_{ij}b^l_{k}-b_{ik}b^l_{j},\quad
\frac{\partial b_{ij}}{\partial x^k}-\frac{\partial b_{ik}}{\partial
    x^j}=\Gamma^s_{ik}b_{is}-\Gamma^s_{ij}b_{ks}
\eqno (12) $$
where $ b^j_i=g^{jl}b_{il} $ and the curvature tenzor has the form
$$ R^l_{ijk} = \frac{\partial \Gamma^l_{ij}}{\partial x^k}-\frac{\partial \Gamma^l_{ik}}
{\partial x^j}+\Gamma^s_{ij} \Gamma^l_{ks}-\Gamma^s_{ik} \Gamma^l_{js}
\eqno (13) $$

Let
$ Z = ( r_{x},  r_{y},  n)^{t}$ . Then
$$ Z_{x} = A  Z,\quad Z_{y} = B  Z \eqno (14) $$
where
$$
A =
\left ( \begin{array}{ccc}
\Gamma^{1}_{11} & \Gamma^{2}_{11} & L \\
\Gamma^{1}_{12} & \Gamma^{2}_{12} & M \\
p_{1}           & p_{2}           & 0
\end{array} \right), \quad
B =
\left ( \begin{array}{ccc}
\Gamma^{1}_{12} & \Gamma^{2}_{12} & M \\
\Gamma^{1}_{22} & \Gamma^{2}_{22} & N \\
q_{1}           & q_{2}           & 0
\end{array} \right) \eqno(15)
$$
Hence we get the new form of the MPCE(12)
$$ A_y - B_x + [A, B] = 0 \eqno (16) $$

Let us introduce the orthogonal trihedral[14]
$$ \vec e_{1} = \frac{\vec r_x}{\surd E}, \,\,\,
\vec e_{2} = \vec n, \,\,\, \vec e_{3} = \vec e_{1} \wedge
\vec e_{2}    \eqno(17) $$

Let $ \vec r_x^2 = E = \pm 1 $ and $ F = 0$.  Then the vectors $
\vec e_{j}$   satisfy equations(3), where
$$k =\frac{L}{2}, \quad
\tau = MG^{-1/2}.     \eqno(18)
$$
So this approach is equivalent to the previous approach.

\begin{center}
{\bf 4. The gauge equivalent counterpart of the Myrzakulov
XXII equation}
\end{center}

Note that the M-XXII$_{q}$ equation(7) is integrable as the
L-equivalent counterpart
of the integrable M-XXII$_{s}$ equation(1). It is means that it must allows
the Lax representation. Let us find it. To this end consider the gauge
transformation
$$ \Phi = g^{-1} \Psi_{1} \eqno(19) $$
If we take  $g$ as $g = \Psi_{1}$ as $\lambda = b$ then after the some calculations
we came to the following Lax representation for the M-XXII$_{q}$ equation(7)
$$
\Psi_{1x} = \{ - i(\lambda^{2} - \frac{pq}{4})\sigma_{3} +
\lambda Q\}\Psi_{1},\quad
\Psi_{1t} = 2\lambda^{2}\Psi_{1y} + (\lambda^{2} B_{2}  +  \lambda B_{1}
+ B_{0})\Psi_{1}  \eqno(20)
$$
with
$$
Q =
\left ( \begin{array}{cc}
0   & q \\
-p  & 0
\end{array} \right) ,
B_{2} = \frac{i}{2}V_{1}\sigma_{3},\quad
B_{1} =  i\sigma_{3}Q_{y} - \frac{1}{2}V_{1}Q,\quad
B_{0} = \frac{1}{4}V_{2} - \frac{i}{8}pqV_{1}.
$$

So between the M-XXII$_{s}$(1) and M-XXII$_{q}$(7) equations
take places the gauge equivalence.

\begin{center}
{\bf 5. On the gauge equivalence between  the M-XXII$_{q}$
equation and the Strachan equation}
\end{center}

Now let us consider the following transformation
$$
q^{\prime} = q\exp(-\frac{i}{2}\partial^{-1}_{x}|q|^{2})   \eqno(21)
$$

Then the new variable $q^{\prime}$ satisfies the Strachan equation[15]
$$ iq^{\prime}_{t} + q^{\prime}_{xy} + i(Vq^{\prime})_x = 0, \quad
 V_x = E(|q^{\prime}|^2)_y. \eqno (22) $$

We see that  the M-XXII$_{q}$ equation(7) and the Strachan equation(22) is
gauge eqivalent to each other. The tranformation (21) induces the
following tranformation of the Jost function$\Psi_{1}$
$$ \Psi_{1} = f^{-1} \Psi_{2} \eqno(23) $$
where
$$
f = \exp(-\frac{i}{4}\partial^{-1}_{x}\mid q\mid^{2}\sigma_{3}) =
\Psi^{-1}_{1}\mid_{\lambda=0}.   \eqno(24)
$$
Then the new Jost function $\Psi_{2} $ satisfies the
following set of equations[15]
$$
\Psi_{2x} = \{ - i\lambda^{2} \sigma_{3} +
\lambda Q^{\prime} \}\Psi_{2},\quad
\Psi_{2t} = 2\lambda^{2}\Psi_{2y} + \{\lambda^{2} B^{\prime}_{2}  +
\lambda B^{\prime}_{1} + B^{\prime}_{0})\}\Psi_{2}  \eqno(25)
$$
with
$$
Q =
\left ( \begin{array}{cc}
0            & q^{\prime} \\
-p^{\prime}  & 0
\end{array} \right) ,
$$
and $B^{\prime}_{j}$ are given in [15,14,13].

\begin{center}
{\bf 6. On some integrals of motion }
\end{center}

As integrable the above presented  equations  allow an infinite number of
integrals of motion. It is interesting that the some important integrals
of motion follow from the geometrical formalism that was presented in section II.
So in 2+1 dimensions we have the following \\
{\bf  Theorema-1:}

The 2+1 dimensional nonlinear evolution  equations admit the following integrals of motion
$$
K_{1} = \int \kappa m_2 dxdy, \quad K_{2} = \int \tau m_2 dxdy \eqno(26)
$$
 In particular for the 2+1 dimensional spin
systems this theorema we can reformulate in the following way\\
{\bf  Theorema-2:}

The 2+1 dimensional spin systems admit the following integrals of motion
$$ K_1 = \int \vec S \cdot (\vec S_{x} \wedge \vec S_{y} )dxdy, \quad
 K_2 = \int \frac {[\vec S \cdot (\vec S_{x}\wedge \vec S_{y})]
[\vec S \cdot (\vec S_{x} \wedge \vec S_{xx})]}
{\mid \vec S_{x}\mid^{\frac{5}{2}}}dxdy. \eqno(27)$$
Note that in the last case $G=\frac{1}{4\pi} K_1$ is the well known
topological charge. The proves of these theoremes are given in[14].

\begin{center}
{\bf 7. A nonisospectral problems }
\end{center}

In contrast with  the 1+1 dimensional case, where  $ \lambda_{t} = 0,$
in our case the spectral parameter $ \lambda_{t} \ne const $
and satisfies the following equation
$$
\lambda_{t} = 2\lambda^{2}\lambda_{y}. \eqno(28)
$$

This equation we can solve  using the following Lax representation
$$ h_x = -i\lambda^{2}\sigma_{3}h, \quad
h_t = 2\lambda^{2}h_y. \eqno(29)
$$
The trivial solution is $\lambda = \lambda_1= const $. To find the other
solution let us consider the following general equation
$$
\lambda_{t} = \kappa \lambda^{n}\lambda_{y} \eqno(30)
$$
where $\kappa = const.$ Let
$$
\lambda_{y} = \sum_{j}b_{j}\lambda^{j},\quad
\lambda_{t} = \sum_{j}d_{j}\lambda^{j}. \eqno(31)
$$
where $b_{j}, d_{j}$ are some functions in general of $y, t$.
In particular we can take
$$
\lambda_{t} = \frac{\lambda}{a - \kappa t},\quad
\lambda_{y} = \frac{\lambda}{y + c}. \eqno(32)
$$
where $a (c)$ is real (complex) constant. Hence follows that solution of equation(30)
has the form
$$
\lambda = \lambda_2 = (\frac{y + c}{a - \kappa t})^{\frac{1}{n}}\eqno(33)
$$
So if $n = 1$ we have
$$
\lambda_2 = \frac{y + c}{a - \kappa t}. \eqno(34)
$$
If $n = 2$ we get
$$
\lambda_2 = (\frac{y + c}{a - \kappa t})^{\frac{1}{2}} \eqno(35)
$$
and so on. In our case the solution of (28) has the form(35) with $\kappa = 2$.
We note the corresponding
solutions of the soliton equations is  called the overlapping or
breaking solutions[17]. In this case soliton equations must be solve with help the non-isospectral
version of the inverse scattering transform(IST) method.
Note that unlike with the 1+1 dimensions, where  $ \lambda_{t} = 0, $
in 2+1 dimensions  we have the following integral of motion for
the spectral parameter
$$ K = \int \lambda dy, \quad K_{t} = 0. \eqno(36)
$$

\begin{center}
{\bf 8.  Conclusion}
\end{center}

To conclude, we have found the L-equivalent counterpart of the
M-XXII equation using the differential geometry of curves and surfaces.
Also the gauge equivalence between this counterpart equation and
the Strachan equation is established. Finally we
present the (1+1)-dimensional reductions of the above considered equations.
We have\\
the M-XXII$_{s}$ equation reduces to the following equation
$$
-iS_t=\frac{1}{2}[S,S_{xx}] + \frac{i}{8b^{2}}\vec S^{2}_{x}S_x-2ib^2 S_{x}, \eqno(37)
$$
the M-XXII$_{q}$ equation reduces to the equation
$$
iq_t + q_{xx} + ipqq_x = 0, \quad
ip_t - p_{xx} + iqpp_x = 0 \eqno(38)
$$
the  Strachan equation reduces to the equation
$$
iq_t + q_{xx} + i(qpq)_x = 0, \quad
ip_t - p_{xx} + i(qpp)_x = 0 \eqno(39) $$
where $p = E\bar q$. Note that
equations (38) and (39) are  gauge equivalent each to other[16].

{\bf Finally we also would like to pose the following questions.

Questions

1. What is the Hirota (bilinear) form of the M-XX$_{s}$(1) and M-XXII$_{q}$
(7) equations?

2. How construct the solutions of these equations?

If you have any results(or information) about these and other
multidimensional spin equations please inform me.}

\end{document}